\documentclass{moriond}

\def\be{\begin{equation}}
\def\ee{\end{equation}}
\def\bea{\begin{eqnarray}}
\def\eea{\end{eqnarray}}

\newcommand{\Photo}{\includegraphics[height=35mm]{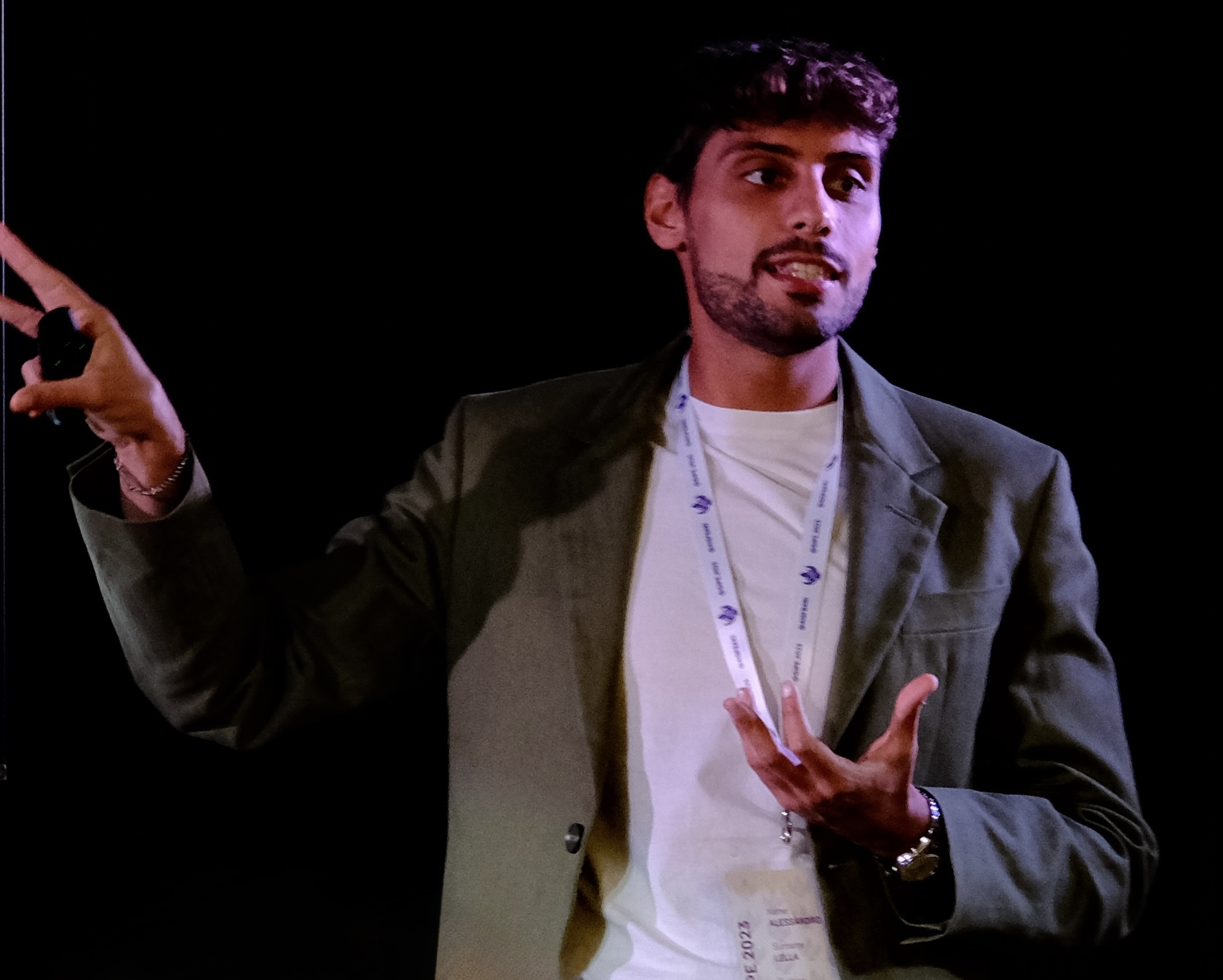}}

\begin{document}
\vspace*{4cm}
\title{Multimessenger probes of Axions from Compact Objects}

\author{Alessandro Lella}

\address{Dipartimento di Fisica e Astronomia, Università degli Studi di Padova,\\Via Marzolo 8, 35131 Padova, Italy\\
Istituto Nazionale di Fisica Nucleare (INFN), Sezione di Padova,\\Via Marzolo 8, 35131 Padova, Italy}

\maketitle\abstracts{
Astrophysics plays a pivotal role in the quest for axions and axion-like particles, offering guidance to experimental efforts and enabling the investigation of axion properties that cannot be probed otherwise. In this context, the extreme conditions in the interiors of compact stellar objects --such as core-collapse supernovae, neutron stars, and binary neutron star mergers-- significantly enhance axion production, providing unparalleled sensitivity to extremely feeble couplings to Standard Model particles. In this context, the techniques of multimessenger astrophysics deepens the understanding of powerful transient events, maximizing the capabilities of current instruments to identify possible signatures of axion emission.}

\section{Introduction}
Several extensions of the Standard Model~(SM) of particle physics predict the existence of novel hidden sectors of new particles interacting extremely weakly with SM. These feebly-interacting particles~(FIPs) are often expected to be light degrees of freedom arising as the low-energy manifestation of some more fundamental UV completions, such as Goldstone bosons of novel global symmetries, or as modes related to the compactification of dark extra-dimensions. The study of these novel sub-GeV particles has received increasing attention over the last 30 years and nowadays constitutes one of the most studied branches of particle physics, commonly referred as the low-energy frontier. Within the vast panorama of possible FIP models, axions and axion-like particles are among the most motivated and widely studied candidates. The QCD axion is a direct prediction of the Peccei-Quinn~(PQ) solution of the strong CP problem of Quantum Chromodynamics~(QCD)~\cite{Peccei:1977hh,Peccei:1977ur}. The problem arises from the presence of an explicit CP-symmetry breaking term in the QCD Lagrangian
\be
    \mathcal{L}_{\rm CP}=\theta\frac{\alpha_s}{8\pi}\tilde{G}_{\mu\nu a}G^{\mu\nu}_{a}
\ee
where $\alpha_s$ is the strong coupling constant, $G_{a}^{\mu\nu}$ is the gluon field strength tensor and $\tilde{G}_{\mu\nu a}=\frac{1}{2}\epsilon_{\mu\nu\rho\sigma}\,G^{\rho\sigma}_a$ its dual, while $\theta$ parametrizes the degree of CP violation . Notably, the $\theta$ term receives contribution from unrelated sectors of the SM, namely the non trivial structure of the QCD vacuum and the quark mass matrix in the weak sector. Many relevant  phenomenological implications arise from the presence of this term, the most notable being its contribution to the neutron electric dipole moment~(nEDM), which has not been observed yet. In this context, the most recent upper limit on the nEDM require $\theta\lesssim10^{-10}$~\cite{Abel:2020pzs}, implying an extreme fine-tuning of the unrelated quantities contributing to $\theta$. In the Peccei-Quinn solution of the strong CP problem, the axion emerges as a pseudo-Nambu-Goldstone boson of a novel global axial symmetry. Then, the CP-violating term can be reabsorbed into the definition of this axion field, promoting $\theta$ from a constant parameter to a dynamical field that is naturally driven to zero by its interaction with QCD. Eventually, the QCD axion acquires a mass as a consequence of the same mechanism, implying that QCD axion mass is expected to be proportional to its couplings with SM particles. In parallel, many others UV completions of the SM, predict additional global axial symmetries whose spontaneous breaking introduces novel pseudoscalar Nambu-Goldston bosons. Since they typically share the same interaction vertices as the QCD axion, these novel pseudoscalar particles are dubbed \emph{axion-like particles}~(ALPs). The main difference between the QCD axion and a generic ALP is in the relation between their mass and couplings. Indeed, ALP couplings are in general unrelated to their masses, so that the ALP mass spectrum spans a wide range of scales from $m_a\lesssim10^{-10}\,$eV up to $m_a\gtrsim1\,$GeV, leading to a rich phenomenology. For the sake of simplicity, in the following both the axion solving the strong CP problem and a generic ALP will be referred to as ``axions'', with explicit distinction made when referring to probes of the QCD axion.

The past 20 years have witnessed a proliferation of many exciting experimental proposals aimed at discover the ALPs and the QCD axion. However, due to their extremely-feeble interactions at low-energies, axion searches require a paradigmatic shift from the \emph{energy frontier}  to the \emph{intensity frontier}, where sensitivity for discovery is driven by a higher particle collisions rather than higher energy thresholds. In this context, stars could provide uniquely powerful laboratories to probe the existence of axions. Such weakly-interacting particles can be copiously produced in the hot ad dense stellar interiors, modifying stellar evolution, giving rise to indirect signatures observable in modern telescopes or producing direct signals in ground-based detectors. A well-known example in which astrophysical searches have demonstrated their complementary role in guiding experimental tests is provided by the case of the famous claim made by the PVLAS experiment. In 2005 the collaboration reported an optical rotation of the polarization direction in a laser beam propagating in an external magnetic field, potentially attributed to an axion~\cite{PVLAS:2005sku}. However, it was quickly pointed out that the axion interpretation would be incompatible with the very existence of the Sun, leading the collaboration to refuse the claim shortly afterwords. 

Motivated by the growing interest in astrophysical axion searches, this contribution aims to provide a non-exhaustive overview of recent results on axion signatures from some of the most promising stellar environments, namely core-collapse Supernovae (SNe), Neutron Stars~(NSs) and Binary Neutron Star~(BNS) mergers.  For further reading, recent reviews on axion astrophysics include Refs.~\cite{Caputo:2024oqc,Carenza:2024ehj,Arza:2026rsl}. The reader is also referred to the GitHub repository~\cite{AxionLimits} for more-detailed information on the limits discussed here.

\section{Impact on the Supernova cooling}

Core-collapse SNe are recognized as one of the most powerful astrophysical factories of axions. The series of catastrophic events leading to the gravitational collapse of massive stars induces extreme conditions in the inner regions of the core, characterized by temperatures $T\sim30-40\,$MeV and densities around the nuclear saturation $\rho\sim10^{14}\,$g cm$^{-3}$. The cooling process of the compact object at the center of the SN core, dubbed Proto-Neutron Star~(PNS), proceeds via the emission of neutrinos of all species, releasing an enormous amount of energy $E_{\rm tot}\sim10^{53}\,$erg over typical time scales in the order $t_{\rm cool}\sim10\,$s. In such extreme environments, the production of axions and ALPs could be significantly enhanced, providing an exceptional astrophysical laboratory to probe these novel elusive particles. 

The most efficient axion production mechanisms in the hot and dense SN core are triggered by their coupling to nucleons $g_{aN}$, with $N=n,\,p$ for neutrons and protons respectively~\cite{Lella:2022uwi,Lella:2023bfb,Lella:2024dmx}. The first considered channel for axion emission from the SN nuclear medium is $NN$ bremsstrahlung $N+N \rightarrow N+N+a$. The computation of the emission rate associated to this process has proven to be highly non-trivial. In this context, it was showed that corrections beyond the usual one-pion exchange (OPE) approximation for the nuclear interaction potential, as well as effective in-medium nucleon masses and multiple nucleon scattering effects may significantly impact axion emissivity~\cite{Carenza:2019pxu}. In particular, the large uncertainties related to the modeling of nuclear interactions at densities close to nuclear saturation --which cannot be driven by experimental data-- suggest that the evaluation of axion emissivity from the SN core should be limited to an order of magnitude estimation, without any claim to precision physics~\cite{Fiorillo:2025gnd}. Furthermore, recent works~\cite{Fore:2019wib} pointed out that strong interactions in the PNS can magnify the abundance of negatively-charged pions, which may reach fractions at the level of $Y_\pi\sim10^{-2}$. Under these conditions, it was realized that the previously overlooked contribution from pionic Compton-like processes $\pi+N \rightarrow a+N$ could be comparable to --or even dominate over-- the $NN$ bremsstrahlung contribution~\cite{Carenza:2020cis}. However, the reader should be warned that the behavior of pionic matter at densities close to nuclear saturation is far to be fully understood, and recent results~\cite{Fore:2023gwv} suggest a reduced pion abundance in the core compared to what originally thought. Hence, pionic processes represents an additional source of uncertainty in the estimation for the SN axion emissivity.

Under the hypothesis of extremely weak nuclear couplings ${g_{aN}\lesssim10^{-8}}$, axions can leave the star unimpeded. In this \emph{free-streaming regime}, the spectrum is characterized by a bimodal shape, since the emission peak associated with the $NN$ bremsstrahlung is located at energies $E_a\sim50\,$MeV, while the pion conversion contribution peaks at $E_a\sim200\,$MeV~\cite{Lella:2022uwi,Lella:2023bfb}. On the other hand, for sufficiently large couplings $g_{aN}\gtrsim10^{-8}$, the SN core becomes ``optically thick'' to the produced axions, which can then be reabsorbed
within the core by means of the inverse processes $N+N+a\rightarrow N+N$ and $a+N\rightarrow\pi+N$~\cite{Lella:2023bfb}. In this regime, dubbed \emph{trapping regime}, axions are not able to escape the inner regions of the core and their emission can be approximately described as a blackbody radiation from a last-scattering surface, in close analogy with the case of neutrinos. Then the axion emission spectrum peaks at energies determined by the temperature of the emission surface around $E_a\sim10\,$MeV~\cite{Lella:2023bfb}.

\begin{figure*}[t!]
\centering
\includegraphics[width=0.7\columnwidth]{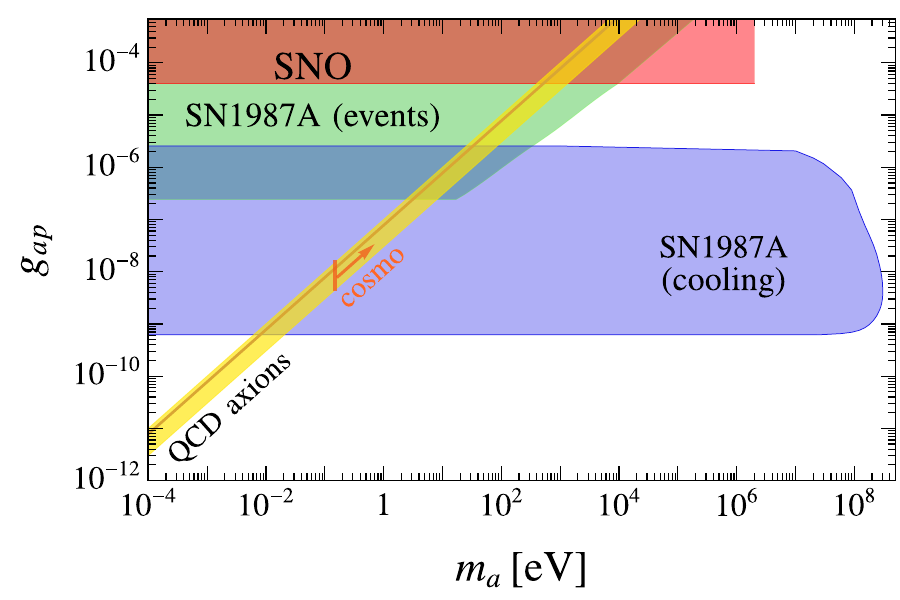}
\caption{Summary plot of the SN 1987A bounds in the $g_{ap}$ vs $m_a$ plane together with the QCD axion band (in yellow). The region labeled SNO is excluded by the search for $p+d \to\,^3 \mathrm{He}+a\,(5.5\,\mathrm{MeV})$ solar axion flux in SNO data. The green and blue regions labeled SN 1987A are ruled out from the non-observation of extra events inside the KII experiment and by the cooling argument, respectively. The orange line with the arrow within the QCD axion band shows the sensitivity of future cosmological probes.}
\label{fig:CoolingBound}
\end{figure*}

Axions can act as a an additional energy-loss channel channel during the SN cooling phase. In particular, if axion emission is too efficient, they might have reduced the energy budget available for neutrino emission, shortening the duration of the neutrino burst observed from the SN 1987A event. Roughly, consistency with observations requires that the axion luminosity $L_a$ computed on the unperturbed model must not exceed the total neutrino luminosity $L_\nu$ provided by the same SN simulation at $1\,$s after the bounce of the SN core. This criterion was employed to exclude the blue region in Fig.~\ref{fig:CoolingBound}~\cite{Lella:2023bfb}, corresponding to axion-proton couplings $5\times10^{-10}\lesssim g_{ap}\lesssim3\times10^{-6}$ for $m_a\lesssim30\,$MeV and $g_{an}=0$. A similar upper bound on the axion-nucleon coupling can be derived from the analysis of the cooling process of cold isolated neutron stars with ages around $\sim10^{5}\,$years~\cite{Buschmann:2021juv}.

SN axions sporting strong couplings to nucleons may lead to a detectable signal in large neutrino water-Cherenkov detectors. In this regard, Engel, Seckel and Hayes proposed to look for possible axion-induced excitation of oxygen nuclei, followed by  the emission of a detectable photon to relax the system {$a+ {^{16}}{\rm O}\rightarrow {^{16}}{\rm O}^* \rightarrow {^{16}}{\rm O}+\gamma$}~\cite{Engel:1990zd}.
The calculation of the cross-section for the axion-oxygen cross section has been recently updated with the usage of the state of the art nuclear models~\cite{Carenza:2023wsm}. The de-excitation spectrum of $^{16}$O$^*$ through $\gamma$-cascades and particle emission, takes into account the emission of nucleons and $\alpha$-particles, as well as the $\gamma$-decays of the final nuclides.
The absence of any evidence for axion-induced events in the detector allows one to rule out values of $g_{ap}$ leading to
\mbox{$N_{\rm ev} \gtrsim 2\,\sqrt{\overline{n}_{\text{bkg}}\Delta t_a}$},
where $\Delta t_a$ is the expected duration of the axion signal and $\overline{n}_{\text{bkg}}\simeq0.02\,$s$^{-1}$ is the average background count rate in Kamiokande II in the time window around SN 1987A. This analysis rules out the green contour in Fig.~\ref{fig:CoolingBound}, excluding the region between the SN 1987A cooling bound and the region excluded by the searches for solar axions in the Sudbury Neutrino Observatory (SNO) data~\cite{Bhusal:2020bvx}. The sensitivity of future cosmological probes to QCD axions is also reported as an orange arrow on the QCD axion band, suggesting that SN bounds make it unlikely that such probes will be able to detect any signature of thermally produced hot dark matter axions. These results underline the power of core-collapse SNe in probing axion phenomenology: even the sparse observations of the SN 1987A neutrino brurst can rule out all the QCD axion masses $m_a\gtrsim10\,$meV, setting one of the most stringent limits on the QCD axion.

\section{Gamma-ray signatures}
In the previous section a minimal model in which axions couple exclusively with nuclear matter was discussed, showing that they may lead to a visible impact on the SN neutrino burst and observable signals in neutrino detectors. On the other hand, if axions feature an additional coupling to photons $g_{a\gamma}$, the range of phenomenological scenarios associated with their emission from compact objects widens significantly. In particular, light axions with masses $m_a\lesssim10^{-3}\,$eV can be efficiently converted into gamma-rays within astrophysical magnetic fields, while MeV-scale axions may decay in photon pairs yielding a flux of MeV-photons inducing various phenomenological consequences. 

\subsection{Light axions}
Once produced in the hottest regions of core-collapse SNe and BNS event, weakly-coupled axions can stream freely out of the stellar volume. Then, they can encounter different kinds of astrophysical magnetic fields along their path from the SN to the Earth. While traveling within a magnetic domain, axions coupled to photons may give rise to the phenomenon of axion-photon oscillations~\cite{Raffelt:1987im}, in which a fraction of the emitted axions is converted into an observable gamma-ray flux. For each magnetic domain with typical size $L$ and strength $B$, it is possible to define an axion-photon oscillation length $L_{\rm osc}$, which typically depends on the properties of the axion field and the surrounding medium. For magnetic fields with strength $B\ll B_{\rm cr}\simeq4.4\times10^{13}\,$G, effects on the axion-photon mixing due to magnetic birefringence can be neglected and the typical oscillation length scales as $L_{\rm osc}\sim m_a^{-2}$. Axion-photon conversions are maximally efficient till the condition $L_{\rm osc}\gg L$ is met and the conversion probability results to be energy independent
\be
    P_{a\gamma}\propto g_{a\gamma}^2B^2L^2\,.
\ee
As the axion mass increases, this condition is not fulfilled anymore and axion-photon oscillations start to loose coherence, resulting in a dramatic reduction in the conversion probability scaling as $P_{a\gamma}\sim m_{a}^{-4}$. Therefore, for each magnetic domain encountered by the emitted axions, it is possible to identify a critical axion mass above which axion-photon conversions become inefficient.

As discussed in the previous section, the high temperatures expected in the core of SNe imply that the axion spectrum has to peak around energies in the order of tens of MeV. As in the coherent mass range the axion-photon conversion probability is energy-independent there is a one-to-one correspondence between the axion energy and the photon energy after conversion.  Therefore, conversions of axions emitted from SN 1987A might have produced a gamma-ray signal at energies $E_a\sim60\,{\rm MeV}$ in coincidence with the observed neutrino burst. The lack of such signal in the Gamma-Ray Spectrometer (GRS) on board of the Solar Maximum Mission (SMM) leads to stringent bounds on the axion parameter space. For axion masses $m_a\lesssim 10^{-9}$~eV, the dominant contribution to the conversion probability is given by the magnetic field hosted by the Milky Way. In particular, measurements of the polarized synchrotron and dust emission suggest typical values of the magnetic field magnitude $B\sim \mathcal{O}(\mu{\rm G})$ with magnetic domains correlated over sizes $L\sim \mathcal{O}(1)$~kpc. In the scenario in which axions were produced in the core of SN 1987A via nuclear couplings and then converted within the Galactic magnetic field, the non-observation of any signal by the SMM rules out values of the product of the coupling $g_{ap}\times g_{a\gamma}\gtrsim4\times10^{-24}\,{\rm GeV}^{-1}$ for $m_a\lesssim10^{-9}\,{\rm eV}$~\cite{Fiorillo:2025gnd} with small variations depending on the choice of the magnetic field model. As the axion mass increases, conversions in the Galaxy become inefficient and axion-photon conversions within the SN progenitor magnetic field may eventually dominate~\cite{Manzari:2024jns}. However, magnetic fields surrounding red/blue supergiant stars are significantly less constrained than the Milky-Way field, with typical surface field strengths within $B_{0}\sim1-100\,{\rm G}$. Furthermore, there is no measurement of the stellar magnetic field available for Sanduleak -69202, the SN 1987A progenitor. Assuming values of the surface field $B_0\sim30-100\,{\rm G}$, SMM observations in coincidence to SN 1987A would probe values of the couplings down to $g_{ap}\times g_{a\gamma}\sim0.4-1.2\times\times10^{-21}\,{\rm GeV}^{-1}$.

Nevertheless, close SN events are extremely rare, occurring at a rate of about 1-3 per century in our Galaxy~\cite{Adams:2013ana}. For this reason, the analysis of gamma-ray bursts from more frequent extragalactic SNe (few per year within a radius of $\sim20\,{\rm Mpc}$) was also employed in previous works to point out possible axion signatures~\cite{Meyer:2020vzy,Crnogorcevic:2021wyj}. Nevertheless, unlike Galactic SNe, the core-collapse time of extra-galactic events cannot be precisely determined as the neutrino bursts from such distances are undetectable with current and near-future underground experiments. In the absence of the neutrino time-trigger identifying the time window of the axion burst, this kind of analysis are dominated by the gamma-ray background. Nevertheless, axion emission from extragalactic SNe has been recently re-analyzed~\cite{Lecce:2025vjc}, highlighting how the nearby environments which may host the SN event (starburst galaxies or Galaxy clusters) may be characterized by magnetic fields more powerful than the Milky-Way, enhancing axion-photon conversions. Furthermore, the advent of future deci-hertz gravitational-wave~(GW) interferometers may allow the detection of the GW signals from core-collapse SNe out to distances of a few tens of Mpc. These instruments would allow the timing of extragalactic events to be determined with unprecedented precision, transforming searches of axion signals from a background-dominated to a background-free analysis. With such time trigger, current and future gamma-ray experiments will be able to probe unexplored regions of the axion parameter space.

Beside SNe, also the compact remnant formed immediately after BNS merger events can reach temperatures comparable to those in a SN core. Therefore this environment would provide an additional extragalactic-source of axions, which may convert both in Galactic fields and in the strongly-magnetized medium($B_0\sim10^{14}\,{\rm G}$) surrounding the remnant. Also in this case, the conversion of such axions could produce a gamma-ray signal coincident with the gravitational-wave emission from the merger event. The sensitivity of \emph{Fermi}-LAT and future proposed gamma-ray detectors to GRBs sourced by axion emission has been investigated in recent studies~\cite{Lecce:2025dbz,Fiorillo:2025gnd}. Nevertheless, the probability of observing an NSM event close enough to probe currently unconstrained regions of the axion parameter space remains extremely low.

\begin{figure}
    \centering
    \includegraphics[width=0.7\columnwidth]{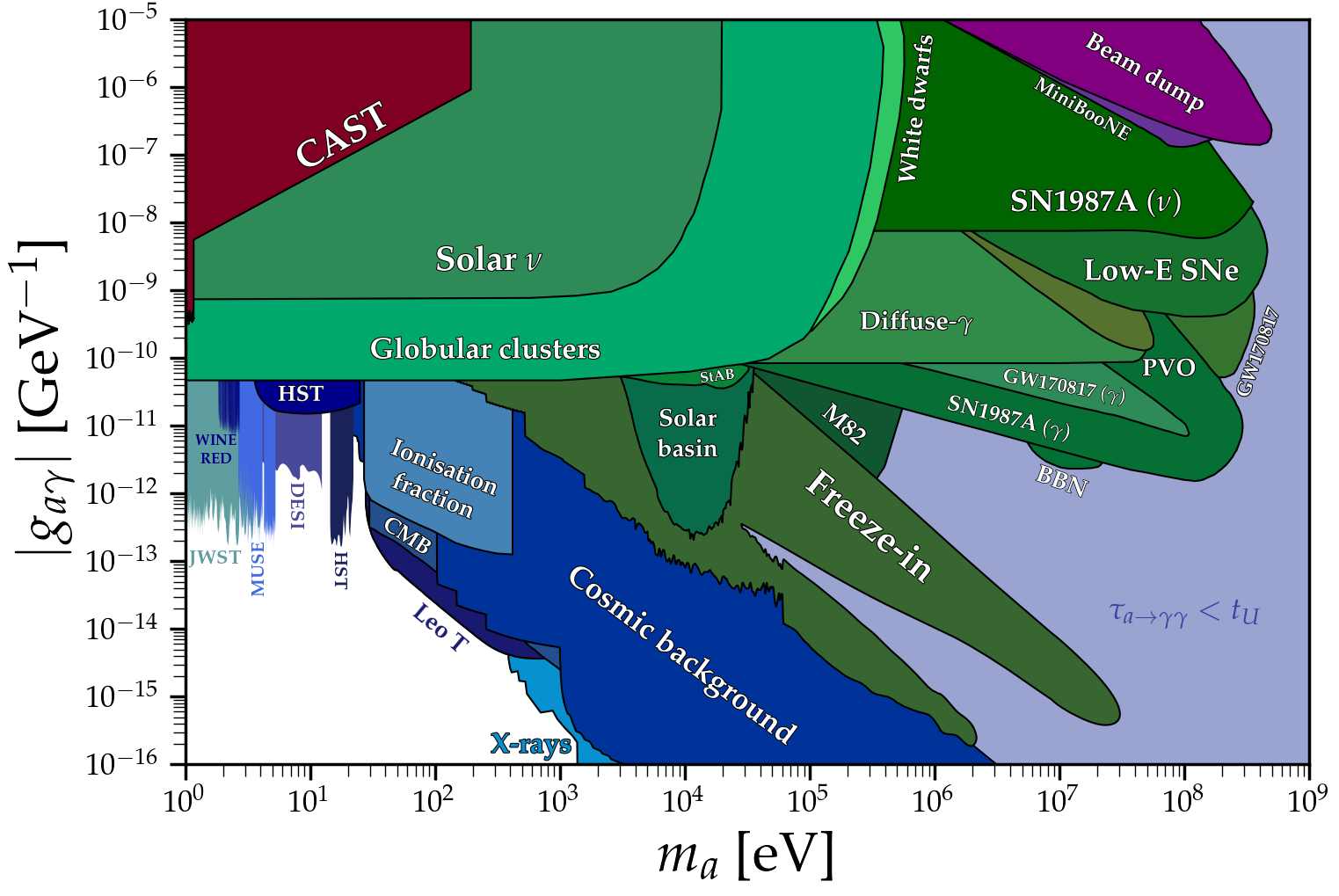}
    \caption{Summary plot of the bounds on the axion--photon coupling for MeV-scale axions, showing astrophysical (green), cosmological (blue), and laboratory (violet) constraints.}
    \label{fig:boundsgag}
\end{figure}

\subsection{MeV-scale axions}

In the opposite mass limit, axions with masses $m_a\gtrsim10\,{\rm MeV}$ emitted from powerful astrophysical environments cannot efficiently convert in photons within astrophysical magnetic fields, but they could decay in photon pairs along their path form the PNS to the Earth. The typical decay length for an axion with energy $E_a$ within the laboratory reference frame is given by
\begin{equation}
        \lambda_{a\gamma\gamma}=\frac{\gamma_a \beta_a}{\Gamma_{a\gamma\gamma}}\simeq 1.3\,{\rm kpc}\left(\frac{\omega_{a}}{100\,{\rm MeV}}\right)\left(\frac{m_{a}}{10\,{\rm MeV}}\right)^{-4}\left(\frac{g_{a\gamma}}{10^{-13}{\rm GeV}^{-1}}\right)^{-2}\sqrt{1-\left(\frac{E_a}{m_a}\right)^2}\,,
\end{equation}
where $\gamma_a$ is the Lorentz factor and $\beta_a$ is the axion velocity. Depending on the value of the axion decay length, the possibility of radiative decays of heavy axions with typical energies $\sim10-250\,{\rm MeV}$ give rise to a variaety of phenomenological scenarios to constrain their parameter space. This section briefly reviews some of them. Fig.~\ref{fig:boundsgag}, taken from the Axion Limits GitHub repository~\cite{AxionLimits}, summarizes current limits on MeV-scale axions. For reference, a minimal model is assumed in which axion production --via photon coalescence $\gamma+\gamma\to a$ and Primakoff process $\gamma+Ze\to a+Ze$-- as well as axion decays are uniquely determined by the axion-photon coupling. 

Axions with masses $m_a\gtrsim100\,{\rm MeV}$ produced in the SN core, may decay within the SN mantle, as their typical decay length is shorter than the SN envelope radius $\lambda_\gamma\lesssim R_{\rm env}$. This mechanism may dump a large amount of energy inside the envelope of the progenitor star. If the energy deposition is too large, axion decays would power the ejection of the outer layers of the stellar mantle independently of any viable SN explosion mechanism. Thus, it is necessary to conservatively require that the axion decays do not deposit an amount of energy larger than the explosion energy observed in past SN events, providing a ``calorimetric'' constrain. In this context, low-energy SNe --a class with particularly low explosion energies-- can severely constrain this scenario~\cite{Caputo:2022mah}. This argument rules out the region labeled ``Low-Energy SNe'' in Fig.~\ref{fig:boundsgag}.

Axions with weaker couplings to photons have typical decay lengths larger than the stellar radius $\lambda_a\gtrsim R_{\rm env}$. Thus, they can leave the star unimpeded and then decay in the interstellar medium between the SN and the Earth. If the SN is sufficiently close, some of the photons produced in radiative decays may eventually reach gamma-ray detectors, which would observe a time-delayed axion-induced gamma-ray burst following the observation of the neutrino burst. Once again, gamma-ray observations of SN 1987A performed by SMM provide a powerful probe of this scenario. In particular, the absence of any detected gamma-ray signal in the operational energy range of this experiment was $[25,100]\,{\rm MeV}$ over a time-window $\Delta t = 223 $~s following the first neutrino signal can be used to rule out the regions of the axion parameter space that would lead to a gamma-ray flux above the detector sensitivity~\cite{Jaeckel:2017tud,Hoof:2022xbe,Muller:2023vjm}. 

Finally, differently for the case of light axions, multimessenger observations of BNS mergers provide a powerful tool to constrain weakly-interacting axions with masses $m_a\sim100-300\,{\rm MeV}$. Axions produced in the hypermassive NS generated by the merger event can decay in photons in the surrounding of the remnant. However, these photons do not propagate freely and may eventually produce a fireball, namely a plasma shell of thermalized photon fluid with temperature $\sim100\,{\rm keV}$. The energies of the resulting photons reaching the Earth would be reprocessed from $\sim100\,{\rm MeV}$ down to $\sim100\,{\rm keV}$~\cite{Diamond:2023cto}. X-ray observations of GW170817 through CALET CGBM, Konus-Wind, and Insight-HXMT/HE, allow to probe the axion-sourced fireball scenario, ruling out regions of the parameter space unconstrained by other arguments. Notice that a similar effect would apply to regions of the axion parameter space probed by SMM searches for axions from SN 1987A. Therefore, these regions would be appropriately constrained by observations from the Pierre Venus Orbiter~(PVO), rather than SMM~\cite{Diamond:2023scc}.

\section{Conclusions}
This contribution aims to highlight the crucial role of astrophysics in the axion quest by providing powerful probes complementary to laboratory searches. In this context, compact objects constitute privileged environments, as the extreme conditions of temperature and density in the stellar interior may significantly enhance axion production. In particular, observations of the cooling process of SN 1987A via neutrino emission have led to stringent constraints on axion properties, and similar limits can be retrieved via the analysis of the cooling process of cold, isolated NSs. The panorama of possible phenomenological scenarios broadens significantly if the axions escaping the cores of compact objects sport a coupling to photons, which enables axion-photon conversions and radiative decays for light and MeV-scale axions, respectively. This discussion underlines that astrophysical searches of axion signatures benefit from the employment of a multimessenger approach, which maximally exploits the capabilities of different experiments in probing cosmic messengers from powerful stellar environments. In the near future, the increasing synergy among  neutrino, gamma-ray and GW searches is expected to provide a decisive imprint in the hunt for the axion. 

\section{Acknowledgments}
The work of Alessandro Lella is supported by the Italian MUR through the FIS 2 project FIS-2023-01577 (DD n. 23314 10-12-2024, CUP C53C24001460001), and by Istituto Nazionale di Fisica Nucleare (INFN) through the Theoretical Astroparticle Physics (TAsP) project.

\section*{References}
\bibliography{moriond}

\begin{thebibliography}{10}

\bibitem{Peccei:1977hh}
R.~D. Peccei and Helen~R. Quinn.
\newblock {\em Phys. Rev. Lett.}, 38:1440--1443, 1977.

\bibitem{Peccei:1977ur}
R.~D. Peccei and Helen~R. Quinn.
\newblock {\em Phys. Rev. D}, 16:1791--1797, 1977.

\bibitem{Abel:2020pzs}
C.~Abel et~al.
\newblock {\em Phys. Rev. Lett.}, 124(8):081803, 2020.

\bibitem{PVLAS:2005sku}
E.~Zavattini et~al.
\newblock {\em Phys. Rev. Lett.}, 96:110406, 2006.
\newblock [Erratum: Phys.Rev.Lett. 99, 129901 (2007)].

\bibitem{Caputo:2024oqc}
Andrea Caputo and Georg Raffelt.
\newblock {\em PoS}, COSMICWISPers:041, 2024.

\bibitem{Carenza:2024ehj}
Pierluca Carenza, Maurizio Giannotti, Jordi Isern, Alessandro Mirizzi, and Oscar Straniero.
\newblock {\em Phys. Rept.}, 1117:1--102, 2025.

\bibitem{Arza:2026rsl}
A.~Arza et~al.
\newblock 3 2026.

\bibitem{AxionLimits}
Ciaran O'Hare.
\newblock cajohare/axionlimits: Axionlimits.
\newblock \url{https://cajohare.github.io/AxionLimits/}, July 2020.

\bibitem{Lella:2022uwi}
Alessandro Lella, Pierluca Carenza, Giuseppe Lucente, Maurizio Giannotti, and Alessandro Mirizzi.
\newblock {\em Phys. Rev. D}, 107(10):103017, 2023.

\bibitem{Lella:2023bfb}
Alessandro Lella, Pierluca Carenza, Giampaolo Co', Giuseppe Lucente, Maurizio Giannotti, Alessandro Mirizzi, and Thomas Rauscher.
\newblock {\em Phys. Rev. D}, 109(2):023001, 2024.

\bibitem{Lella:2024dmx}
Alessandro Lella, Eike Ravensburg, Pierluca Carenza, and M.~C.~David Marsh.
\newblock {\em Phys. Rev. D}, 110(4):043019, 2024.

\bibitem{Carenza:2019pxu}
Pierluca Carenza, Tobias Fischer, Maurizio Giannotti, Gang Guo, Gabriel Mart{\'\i}nez-Pinedo, and Alessandro Mirizzi.
\newblock {\em JCAP}, 10(10):016, 2019.
\newblock [Erratum: JCAP 05, E01 (2020)].

\bibitem{Fiorillo:2025gnd}
Damiano F.~G. Fiorillo, {\'A}ngel Gil~Muyor, Hans-Thomas Janka, Georg~G. Raffelt, and Edoardo Vitagliano.
\newblock {\em JCAP}, 03:053, 2026.

\bibitem{Fore:2019wib}
Bryce Fore and Sanjay Reddy.
\newblock {\em Phys. Rev. C}, 101(3):035809, 2020.

\bibitem{Carenza:2020cis}
Pierluca Carenza, Bryce Fore, Maurizio Giannotti, Alessandro Mirizzi, and Sanjay Reddy.
\newblock {\em Phys. Rev. Lett.}, 126(7):071102, 2021.

\bibitem{Fore:2023gwv}
Bryce Fore, Norbert Kaiser, Sanjay Reddy, and Neill~C. Warrington.
\newblock {\em Phys. Rev. C}, 110(2):025803, 2024.

\bibitem{Buschmann:2021juv}
Malte Buschmann, Christopher Dessert, Joshua~W. Foster, Andrew~J. Long, and Benjamin~R. Safdi.
\newblock {\em Phys. Rev. Lett.}, 128(9):091102, 2022.

\bibitem{Engel:1990zd}
J.~Engel, D.~Seckel, and A.~C. Hayes.
\newblock {\em Phys. Rev. Lett.}, 65:960--963, 1990.

\bibitem{Carenza:2023wsm}
Pierluca Carenza, Giampaolo Co, Maurizio Giannotti, Alessandro Lella, Giuseppe Lucente, Alessandro Mirizzi, and Thomas Rauscher.
\newblock {\em Phys. Rev. C}, 109(1):015501, 2024.

\bibitem{Bhusal:2020bvx}
Aagaman Bhusal, Nick Houston, and Tianjun Li.
\newblock {\em Phys. Rev. Lett.}, 126(9):091601, 2021.

\bibitem{Raffelt:1987im}
Georg Raffelt and Leo Stodolsky.
\newblock {\em Phys. Rev. D}, 37:1237, 1988.

\bibitem{Manzari:2024jns}
Claudio~Andrea Manzari, Yujin Park, Benjamin~R. Safdi, and Inbar Savoray.
\newblock {\em Phys. Rev. Lett.}, 133(21):211002, 2024.

\bibitem{Adams:2013ana}
Scott~M. Adams, C.~S. Kochanek, John~F. Beacom, Mark~R. Vagins, and K.~Z. Stanek.
\newblock {\em Astrophys. J.}, 778:164, 2013.

\bibitem{Meyer:2020vzy}
Manuel Meyer and Tanja Petrushevska.
\newblock {\em Phys. Rev. Lett.}, 124(23):231101, 2020.
\newblock [Erratum: Phys.Rev.Lett. 125, 119901 (2020)].

\bibitem{Crnogorcevic:2021wyj}
Milena Crnogor{\v{c}}evi{\'c}, Regina Caputo, Manuel Meyer, Nicola Omodei, and Michael Gustafsson.
\newblock {\em Phys. Rev. D}, 104(10):103001, 2021.

\bibitem{Lecce:2025vjc}
Francesca Lecce, Alessandro Lella, Giuseppe Lucente, Maurizio Giannotti, and Alessandro Mirizzi.
\newblock 12 2025.

\bibitem{Lecce:2025dbz}
Francesca Lecce, Alessandro Lella, Giuseppe Lucente, Vimal Vijayan, Andreas Bauswein, Maurizio Giannotti, and Alessandro Mirizzi.
\newblock {\em Phys. Rev. D}, 112(2):023001, 2025.

\bibitem{Caputo:2022mah}
Andrea Caputo, Hans-Thomas Janka, Georg Raffelt, and Edoardo Vitagliano.
\newblock {\em Phys. Rev. Lett.}, 128(22):221103, 2022.

\bibitem{Jaeckel:2017tud}
J.~Jaeckel, P.~C. Malta, and J.~Redondo.
\newblock {Decay photons from the axionlike particles burst of type II supernovae}.
\newblock {\em Phys. Rev. D}, 98(5):055032, 2018.

\bibitem{Hoof:2022xbe}
Sebastian Hoof and Lena Schulz.
\newblock {\em JCAP}, 03:054, 2023.

\bibitem{Muller:2023vjm}
Eike M{\"u}ller, Francesca Calore, Pierluca Carenza, Christopher Eckner, and M.~C.~David Marsh.
\newblock {\em JCAP}, 07:056, 2023.

\bibitem{Diamond:2023cto}
Melissa Diamond, Damiano F.~G. Fiorillo, Gustavo Marques-Tavares, Irene Tamborra, and Edoardo Vitagliano.
\newblock {\em Phys. Rev. Lett.}, 132(10):101004, 2024.

\bibitem{Diamond:2023scc}
Melissa Diamond, Damiano F.~G. Fiorillo, Gustavo Marques-Tavares, and Edoardo Vitagliano.
\newblock {\em Phys. Rev. D}, 107(10):103029, 2023.
\newblock [Erratum: Phys.Rev.D 108, 049902 (2023)].

\end{thebibliography}

\end{document}